\RequirePackage{fix-cm}
\documentclass[smallcondensed]{svjour3}     
\smartqed  
\usepackage{graphicx}
\usepackage{latexsym}
\usepackage{textcmds}
\usepackage{float}
\usepackage{comment}
\usepackage{graphicx}
\usepackage{booktabs}
\usepackage{xcolor}
\usepackage{multirow}
\usepackage{dirtree,varwidth}
\usepackage{appendix}
\usepackage{enumitem}

\usepackage{tikz}

\usepackage{array}
\newcolumntype{L}[1]{>{\raggedright\let\newline\\\arraybackslash\hspace{0pt}}m{#1}}
\newcolumntype{C}[1]{>{\centering\let\newline\\\arraybackslash\hspace{0pt}}m{#1}}
\newcolumntype{R}[1]{>{\raggedleft\let\newline\\\arraybackslash\hspace{0pt}}m{#1}}

\begin{document}

\title{Sports Recommender Systems: Overview and Research Issues
}

\author{Alexander Felfernig  \and Manfred Wundara \and Thi Ngoc Trang Tran \and Viet-Man Le \and Sebastian Lubos \and Seda Polat-Erdeniz}




\institute{Alexander Felfernig \and Thi Ngoc Trang Tran \and Viet-Man Le \and Sebastian Lubos \and Seda Polat-Erdeniz \at
              Institute of Software Technology (Graz University of Technology) - Graz, Austria \\
              \email{\{afelfern,ttrang,vietman.le,slubos,spolater\}@ist.tugraz.at}           
           \and
           Manfred Wundara \at
              Magistrat Villach, Austria\\
              \email{manfred.wundara@villach.at}
}

 \date{Received: date / Accepted: date}

\maketitle

\begin{abstract}
Sports recommender systems receive an increasing attention due to their potential of fostering healthy living, improving personal well-being, and increasing performances in sport. These systems support people in sports, for example, by the recommendation of healthy and performance-boosting food items, the recommendation of training practices, talent and team recommendation, and the recommendation of specific tactics in competitions. 
 With applications in the virtual world, for example, the recommendation of maps or opponents in e-sports, these systems already transcend conventional sports scenarios where physical presence is needed. On the basis of different working examples, we present an overview of sports recommender systems applications and techniques. Overall, we analyze the related state-of-the-art and discuss open research issues.

\end{abstract}


\section{Introduction}\label{sec:introduction}

Recommender systems support users in the identification of items from a potentially complex and large catalog \cite{Alhijawietal2023,BurkeFelfernigGoeker2011,Jannach2010}. These systems are in wide-spread use in various application domains ranging from online purchasing platforms such as amazon.com \cite{SmithLinden2017}, video streaming platforms such as netflix.com \cite{Gomez2016} and youtube.com \cite{Davidson2010}, to different applications in the context of high-involvement items such as houses \cite{Dalyetal2014} and financial services \cite{Felfernig2006}. Due to the emergence of new systems such as wearables,  recommender systems  nowadays do not not only focus on the "online world", but also include various new applications in the "real world" comprising, for example, working environments, recreation, and leisure \cite{Smyth2019}. 

As a result of human evolution, our bodies are designed for physical hunting and gathering activities \cite{Wallaceetal2018} which have a low importance in modern society -- this leads to various civilization diseases \cite{Smyth2019}. Aging societies and increased awareness of the importance of active lifestyles increases the demand of recommendation systems helping people in sports-related contexts. Numerous applications of sports recommenders already exist, ranging from the recommendation  of healthy food  \cite{AlcarazHerrera2022} and the recommendation of training plans \cite{Smyth2022} to the recommendation of adventure playgrounds in e-sports \cite{Wuetal2017}. In this article, we provide an overview of the  state-of-the-art in applying recommender systems in sports-related scenarios -- we denote such systems as \emph{sports recommender systems}.  

Compared to related overviews \cite{Bealetal2019,Chmait2021,Hammes2022}, we provide an in-depth analysis of the application of recommender systems in sports. We include examples with the goal to give insights of how to use the discussed recommendation approach. In our overview, we do not focus on specific sports segments (e.g., elite sports or team sports) but rather provide a more general overview  of sports recommender systems also including aspects such as sports recommendations for elderly persons and persons with diseases. For demonstration purposes and for reasons of understandability, we use basic recommendation approaches, i.e., we do not focus on a detailed discussion of alternative algorithmic solutions.

Insights of our literature analysis are the following: (1) most  approaches in sports do not support recommendations of more complex items such as bundles combining, for example, food recommendations with recommendations for training sessions and training partners. (2) Although highly relevant, the explanation of recommendations still plays a minor role, i.e., explanations do not focus on specific explanation goals in a systematic fashion. (3) The integration of psychological aspects  of persuasion and human decision making \cite{Jameson2015,Kaya2014} is only discussed on the level of individual research works but their integration into real-world systems is the exception of the rule.

The contributions of this article are the following: (1) we provide an in-depth overview of existing recommendation approaches in sports. (2) we give  examples of how to apply recommendation algorithms in the discussed application contexts. (3) based on our analysis of the existing state-of-the-art, we indicate different directions of future research specifically pointing out new applications beyond the existing mainstream of recommender systems research. In this context, we also analyze related potentials of large language models (LLMs).

The remainder of this article is organized as follows. In Section \ref{sec:basicrecommendations}, we provide an overview of basic recommendation approaches. In Section \ref{sec:methodology}, we present the methodology used for our literature analysis.    Thereafter, in Section \ref{sec:sportsrecommendersystems}, we summarize the  state-of-the-art in sports recommender systems. A discussion of  open research issues in provided in Section \ref{sec:researchissues}. The article is  concluded with Section \ref{conclusions}.

\section{Basic Recommendation Approaches}\label{sec:basicrecommendations}
There is no \emph{one size fits all} recommendation approach, but rather different approaches that are applicable depending on the available data sources (e.g., user purchasing behavior or user news reading preferences). To provide an optimal user experience, these systems apply various Artificial Intelligence (AI) techniques such as \emph{explanations}, \emph{machine learning}, and \emph{intelligent user interfaces}. The used data sources and properties of those approaches are summarized in Tables \ref{tab:recommendationapproachesknowledgesources} and \ref{tab:recommendationapproachescomparison}.

\subsection{Collaborative Filtering}
Collaborative filtering (CF) \cite{Ekstrand2011CFBook}  is based on the idea of word-of-mouth promotion by recommending preferred items of nearest neighbors (NNs) which represent users with preferences similar to the current user (e.g., running shoes already purchased and evaluated positively by nearest neighbors). CF allows for an easy setup due to the fact that no detailed information about the offered items is needed. It can help to generate serendipity effects which create a kind of positive "surprise" in terms of completely new items the current user is really happy about. CF is a one-shot recommendation approach which does not require any preference elicitation dialog with the user. A major disadvantage of CF is the so-called \emph{ramp-up} problem meaning that user $\times$ item ratings (of the current user and the nearest neighbors) need to be available before reasonable recommendations can be determined.

\subsection{Content-based Filtering}
Content-based filtering (CBF) \cite{Pazzani2007} recommends items which are similar to those the current user has already consumed in the past. CBF does not take into account the preferences of nearest neighbors which limits the ramp-up problem somehow (only user $\times$ item ratings need to be available but no ratings of nearest neighbors). In contrast to CF, CBF needs some kind of item knowledge, for example, in terms of \emph{keywords} or \emph{categories}. This item knowledge is needed for constructing the user preference profile used for the recommendation of new items (based on the similarity of user profile entries and item descriptions). For example, since a user preferred to purchase running t-shirts of a specific size and color in the past, similar t-shirts will be recommended in future purchasing sessions. Similar to CF, also CBF is a "one-shot" recommendation approach, i.e., not dialog-based. Serendipity effects with CBF are very limited, since  items are recommended which are similar to those already consumed in the past. CBF recommender setup is easy since recommendations are just based on the item preferences of the current user. A major disadvantage of both, CF and CBF, is the limited applicability in  scenarios with a low purchase frequency and  related potential preference shifts.

\subsection{Knowledge-based Recommendation}
Knowledge-based recommender (KBR) systems \cite{Burke2000,FelfernigBurke2008constraintbased} are based on a semantic description of items and user preferences. Since recommendation knowledge is formulated explicitly (either in terms of constraints or similarity metrics), these approaches do not have ramp-up issues. Knowledge-based recommenders are conversational \cite{ChristakopoulouConversational2016}, i.e., user preferences are elicited within a recommendation dialog. The primary types of knowledge-based recommendation are \cite{Aggarwal2016} (1) \emph{constraint-based recommendation} (CON)\cite{FelfernigBurke2008constraintbased} and (2) \emph{case-based recommendation} (CAS) \cite{LorenziRicci2005}. The former determines recommendations on the basis of constraints, the latter recommends items similar to the user preferences. KBRs are applied in complex item domains where  items are described in terms of attributes (properties). KBR systems efficiently handle "no solution could be found" situations by indicating potential adaptations of user requirements in such a way that a solution (recommendation) can be identified \cite{FelfernigetalFastDiag2012}. Setup problems exist in the sense that, for example, specific user dialogs have to be designed and implemented and the recommendation knowledge has to be defined  \cite{Ulz2017}. Finally, serendipity is limited by the encoding of the recommendation knowledge.

\begin{table}[ht]
\centering \caption{Collaborative filtering (CF), content-based filtering (CBF), and knowledge-based recommendation (KBR) with relevant data sources (preferences of the current user, preference history of the current user, nearest neighbor (NN) preferences, and item knowledge).}
\begin{tabular}{|c|c|c|c|c|c|c|c|c|c|c|c|c|c|c|c|c|c|c|} 
\hline
  approach                    &    current preferences           & past preferences & NN preferences   &  item knowledge    \tabularnewline  \hline \hline
  CF          &  -                       &     x              &   x                 &     -              \tabularnewline  \hline
  CBF         &  -                       &     x              &   -                 &     x              \tabularnewline  \hline
  KBR         &  x                       &     -              &   -                 &     x              \tabularnewline  \hline
\end{tabular} 
\label{tab:recommendationapproachesknowledgesources} 
\end{table}

\begin{table}[ht]
\centering \caption{Basic properties of different recommendation algorithms (CF = collaborative filtering, CBR = content-based filtering, and KBR=knowledge-based recommendation).}
\begin{tabular}{|c|c|c|c|c|c|c|c|c|c|c|c|c|c|c|c|c|c|c|} 
\hline
  recommendation approach     &  CF           & CBF & KBR \tabularnewline  \hline \hline
  easy setup                  &  x                        &     x          &   -               \tabularnewline  \hline
  dialog-based                &  -                        &     -          &   x               \tabularnewline  \hline
  serendipity                 &  x                        &     -          &   -               \tabularnewline  \hline
  ramp-up problem             &  x                        &     x          &   -               \tabularnewline  \hline
  high-involvement items      &  -                        &     -          &   x               \tabularnewline  \hline
\end{tabular} 
\label{tab:recommendationapproachescomparison} 
\end{table}

\subsection{Further Recommendation Approaches}
To exploit synergy effects, the recommendation approaches of CF, CBF, and KBR can be systematically integrated into so-called hybrid recommender systems \cite{Burke2002}. For example, CF and CBF can be combined in such a way that in the first step only CBF is activated to collect corresponding user $\times$ item ratings. Having this information available, CF can be activated as well. Finally, \emph{group recommender systems} \cite{masthoff2015group,McCarthyetalSkiGroupRec2006} follow the idea of determining recommendations not for individual users but for whole groups. For example, a group recommender system can be applied to recommend training plans for a football team in such a way that the training needs of individual team members are taken into account as much as possible. Furthermore, group recommenders can propose items that take into account as much as possible  the individual food preferences of team members.

\section{Research Methodology}\label{sec:methodology}

Our overview of different approaches in sports recommender systems is based on a literature analysis conduced following the phases of \emph{searching} (querying of leading research portals and search for publications in related conferences and journals), \emph{reviewing} (evaluation and classification of the identified scientific contributions), and \emph{detailed discussion of evaluated contributions} (deciding which contributions should be included based on the criteria  quality and topic-wise relevance). We have performed queries on the research platforms  Google Scholar\footnote{https://scholar.google.com/}, ResearchGate\footnote{https://www.researchgate.net/}, ScienceDirect\footnote{https://www.sciencedirect.com/}, SpringerLink\footnote{https://link.springer.com/}, and Elsevier\footnote{https://www.elsevier.com/} on the basis of the search keywords "artificial intelligence", “recommender systems”, "sports",  "machine learning", "decision support", “optimization” also taking into account  related combinations. On the basis of the received query results, we have extended the set of search criteria by taking into account additional terms ("deep learning", "data mining", "large language models", "e-sports", and "healthy living"). In addition to the mentioned research platforms, we have analyzed  contributions in topic-relevant conferences and journals including the International Joint Conference on Artificial Intelligence (IJCAI), the AAAI Conference on Artificial Intelligence, the European Conference on Artificial Intelligence (ECAI), the Recommender Systems Conference (RecSys), the User Modeling, Adaptation, and Personalization (UMAP) conference, Artificial Intelligence in Sports (AI-Sports) Workshop, the Constraint Programming (CP) conference, User Modeling and User-Adapted Interaction (UMUAI), ACM Transactions on Recommender Systems, Journal of Sports Science and Medicine (JSM), Journal of Medical Internet Research (JMIR), Journal of Sports Analytics, and Frontiers in Psychology. Based on the snowballing technique \cite{Wohlin2014}, we have analyzed and selected further topic-relevant scientific contributions. The major criterion for a contribution to be included in our overview was high-quality and a clear topic-wise match. In total, we have regarded 140 publications as relevant for this overview.

\begin{table}[!]
\centering  
\caption{Overview of existing  applications of sports recommender systems organized following the recommendation approaches of collaborative filtering (CF), content-based filtering (CBF), constraint-based recommendation (CON), case-based recommendation (CAS), and other (further) recommendation approaches (OTR).} 
\begin{tabular}{|C{1.75cm}|C{1.75cm}| C{1.1cm}|C{1.1cm}|C{1.1cm}|C{1.1cm}|C{1.1cm}|} 
\hline
   application  &  what is        &   \multicolumn{5}{C{5.0cm}|}{recommendation approach} \tabularnewline \cline{3-7} 
     scenario   &  recommended?   & CF  & CBF & CON & CAS & OTR \tabularnewline\hline \hline
                & training plans and activities   & \cite{Deepaketal2022,Heetal2014} & \cite{LeietalOutdoorSportsRecommender2022} & \cite{SantosGagoetal2019RecommenderSportswomenConstraintBasedMenst} & \cite{Berndsenetal2020,Donciuetal2011,FeelyetalExplainableRaceTimePredictions2020,Feely2020,Feelyetal2023ModelingTrainingPractices,Smyth2022} & \cite{Deepaketal2022}\tabularnewline \cline{2-7} 
                & focused trainings  & \cite{Portazetal2023} & - & \cite{Kashino2018Brain,RoanesLozanoetal2020} & - & -\tabularnewline \cline{2-7}
recommendation of trainings                & sports equipment  & - & - & \cite{LoChangSheuJung2014} & - & -\tabularnewline \cline{2-7}
                & videos & \cite{Ezin2018FitnessTF,Odden2017,SunTangSportsDanceVideo2022} & - & - & - & -\tabularnewline \cline{2-7}
                & destinations \& routes   & \cite{Avesanietal2005} & \cite{Ivanovaetal2022,IvanovaIntroducingContext2023,IvanovaWald2023,Lietal2021,Othmanetal2023,Takamaetal2015,Teslyuketal2019} & - & \cite{LoeppZiegler2020,Serdouketal2021} & \cite{McCarthyetalSkiGroupRec2006,Wirzetal2011} \tabularnewline \cline{2-7}
                & e-sports related settings & \cite{Wuetal2017} & - & - & - & - \tabularnewline \hline
                & tactics in endurance sports & - & - & \cite{Berndsenetal2019,SmythCunningham2017,Smyth2022} & - & -  \tabularnewline \cline{2-7} 
tactical \& strategic planning                & tactics against opponents  & \cite{AbreuetalCF2014,Abreuetal2014,Mengetal2020,Sunetal2020TableTennisTactics} & \cite{TsaietalTacticRecognition2017,WuetalVideoActionRecognition2022} & \cite{ChangYu2022TennisTactics} & - & -  \tabularnewline \cline{2-7} 
                & e-Coaching   & - & - & \cite{BorattoetalPredictingWorkoutQuality2018,BorattoetalECoachingEcoSystem2017} & - & \cite{Borattoetal2022eCoaching}  \tabularnewline \hline
recommending individuals \& teams & individuals   & \cite{Blaszczyketal2023,Chenetal2018HeroRecommendation,Klancaretal2019} & \cite{IvanovaBehavior2021} & \cite{Papicetal2009TalentIdentification,Rajeshetal2022} & \cite{YIlmazetal2022} & - \tabularnewline \cline{2-7}
                                  & teams  & \cite{Patil2020,Tuetal2017} & - & \cite{LandersDuperouzel2019,Thavamunietal2023TeamConfiguration} & \cite{Jayanthetal2018} & - \tabularnewline \hline
handling injury \& illness & measures for injury/illness avoidance  & \cite{Berkovskyetal2010,Bhimavarapuetal2021} & \cite{CoppensetalMotivationMove2023,RanaetalContentBasedHealthRecommenderSystems2020} & \cite{DobricanZampuieris2016RehabilitationConstraintBasedRiskAvoidance,Matosetal2020,SantosGagoetal2019RecommenderSportswomenConstraintBasedMenst,SasakiTakama2013} & \cite{Arciniegaetal2023,Feelyetal2021} & -\tabularnewline \cline{2-7} 
                & recovery actions  & \cite{Ismailetal2021,Shambourteal2023} & \cite{Emrichetal2014} & \cite{Yangetal2018} & - & -\tabularnewline \hline
fostering  & nutrition  & \cite{Shrimaletal2021} & \cite{Espinetal2016} & \cite{Yangetal2017} & - & \cite{Vandeputteetal2022} \tabularnewline \cline{2-7} 
 healthy living               & training practices  & \cite{AlcarazHerrera2022} & \cite{Khwajaetal2019,Nietal2019} & \cite{JuliantBaizalDharayani2023,Tsengetal2015DietAndExerciseGuideline} & - & \cite{Tragosetal2023,YomTovetalEcouragingPhysicalActivityDiabetes2017,Zhaoetal2020} \tabularnewline \hline
                & sports events  \cite{Achilleosetal2021,Nguyenetal2019} & \cite{Achilleosetal2021} & \cite{Nguyenetal2019} & - & - & - \tabularnewline \cline{2-7} 
event handling \& recommendation                & event designs  & - & - & \cite{ChmaitetalTennisSuperstars2020} & - & - \tabularnewline \cline{2-7} 
                & event-related items \& games     & \cite{Lenhart2016CombiningCA} & \cite{Pichletal2018SportsAwards,Sanchezetal2012,Subramaniyaswamy2018} & \cite{Pessemieretal2018} & - & - \tabularnewline \hline
\end{tabular} 
\label{tab:recommendergoals} 
\end{table}

\section{Sports Recommender Systems}\label{sec:sportsrecommendersystems}

\emph{Sports recommender systems} focus on the recommendation of sports-related items such as training plans, activities, and event-related information thus supporting people in improving and enhancing their sports-related capabilities and experiences. An overview of  basic application scenarios of sports recommender systems is provided in Table \ref{tab:recommendergoals} -- this categorization is used for structuring our follow-up discussions of different scenarios including the recommendation of: (1) \emph{trainings}  which help athletes and teams to perform at their best  with their developed skills, cognitive capabilities, and physical condition (Section \ref{subsec:trainingpractices}), (2)  \emph{tactics \& strategies}  which are crucial to take optimal decisions leading to success, for example, winning a soccer match or the whole championships (Section \ref{subsec:tacticalplanning}), (3) \emph{individuals and teams}  who have the potential of becoming competitive and successful in a particular sports discipline (Section \ref{subsec:teamconfiguration}), (4) \emph{measures} helping to prevent illness, athlete overloading, and  injuries  and support efficient recovery and regain of a competitive performance level (Section \ref{subsec:injuryhandling}), (5) \emph{practices} regarding aspects such as adequate nutrition, hydration, and stress avoidance which help to assure an expected performance level and personal well-being (Section \ref{subsec:healthyliving}), and (6) \emph{events} and related items such as event information and merchandises helping to attract audiences, increase turnovers, and maximize enjoyment (Section \ref{subsec:eventhandling}). We also analyze the role of \emph{explanations} that can help to foster the setting and sustainable pursuing of personal goals   (Section \ref{subsec:motivation}).

\subsection{Recommendation of Trainings}\label{subsec:trainingpractices}

With \emph{recommendation of trainings} we refer to recommendations that help persons to take care of their sports performance. Related dimensions are the recommendation of \emph{training plans} (e.g., in the context of the preparation for a marathon run), \emph{focused trainings} to improve specific techniques (e.g., to improve the personal tennis service performance but also to learn to better keep the focus in competitions), appropriate \emph{sports equipment} as a major precondition of being competitive, instructive \emph{videos} helping, for example, to better visualize specific exercises, and \emph{destinations and routes} to assure an optimal and enjoyable training environment. In e-sports, related recommendations help to identify challenging playgrounds and opponents as a major precondition to optimally prepare for a competitive setting. 

\subsubsection{Training Plans and Activities}

Recommender systems for training plans are discussed, for example, in Donciu et al. \cite{Donciuetal2011} who introduce a case-based recommendation approach that combines workouts with corresponding diets thus following a basic item bundling approach. He et al. \cite{Heetal2014} propose a physical activity context-aware collaborative filtering based recommendation approach taking into account contextual factors such as  a users risk tolerance, available budget, and weather conditions. A related approach is presented in Santos-Gago et al. \cite{SantosGagoetal2019RecommenderSportswomenConstraintBasedMenst}  who introduce a constraint-based recommendation approach taking into account the current physical situation of athletes and how this can impact physical activities in the context of proposed training plans. To propose different types of outdoor sports activities, Lei et al. \cite{LeietalOutdoorSportsRecommender2022}  discuss a collaborative filtering based (combined with a content-based) recommendation approach (to resolve the cold start problem).

 Case-based recommendation (CBR) can be applied, for example, to infer training practices of persons with a similar running performance.  Approaches to the CBR of training sessions for marathon runners are presented in Berndsen et al. \cite{Berndsenetal2020} (training plans) and Feely et al. \cite{FeelyetalExplainableRaceTimePredictions2020,Feely2020} (training plans and race time predictions). Follow-up works focus on the "inclusion" of recreational marathon runners, for example, in terms of realistic and reasonable training plans \cite{Feelyetal2023ModelingTrainingPractices,Smyth2022}.

\begin{example}[Recommendation of Marathon Pacings] 
A simplified example of marathon training session recommendation is depicted in Table \ref{tab:casebasedpacerecommendation}.  In this example, the case base includes the pacing history of different training sessions of marathon runners and the case-based recommender should identify reasonable training paces for the "current" runner for week$_2$ -- week$_4$. The "case" most similar to the current case is runner$_1$ whose pacings for week$_2$ -- week$_4$ can be recommended.
\end{example}

\begin{table}[ht]
\centering \caption{Recommendation of marathon pacings (per kilometre) using case-based recommendation (CBR) based on example training pacings of successful marathon runners ($p_i$ denotes pacing $i$). For example, the \emph{current} runner should plan a pacing of $6.75$ for the first training session in the second week (following the practices of the nearest "case" of runner 1). Values in boldface represent  recommendations.}
\begin{tabular}{|c|c|c|c|c|c|c|c|c|c|c|c|c|c|c|c|c|c|c|} 
\hline
  week & \multicolumn{2}{c|}{week$_1$}  & \multicolumn{2}{c|}{week$_2$}  & \multicolumn{2}{c|}{week$_3$}  & \multicolumn{2}{c|}{week$_4$}    \tabularnewline \cline{2-9} 
           & $p_1$  & $p_2$   & $p_1$ & $p_2$   & $p_1$  & $p_2$  & $p_1$  & $p_2$ \tabularnewline \hline \hline
  runner$_1$  & 7.75 & 7.5   & \bf 6.75  & \bf6.5     & \bf6.75   & \bf6.5    & \bf6.5  & \bf6.25 \tabularnewline \hline
  runner$_2$  & 8.5  & 8.0  & 8.5   & 8.0    & 7.5    & 7.25   & 7.0  &  6.75 \tabularnewline \hline \hline
  current runner &   8.0  & 7.5  &  ?  & ?    & ?    & ?   & ?  & ? \tabularnewline \hline
\end{tabular} 
\label{tab:casebasedpacerecommendation} 
\end{table}

\begin{example}[Recommendation of Training Partners]
A precondition for sports activities is often the identification of training partners \cite{Tuetal2017}. In tennis training sessions, there is a need of recommendations regarding which team members (players $pl_i$) should act as training partners in  weekly  sessions (see Table \ref{tab:grouprecommendationtennispractices}). 
 An example criteria is that the expertise level difference (measured in so-called $itn$ points between individual player pair) should be minimized and that in each week (training session), the training partner assignment should differ (with the goal in mind to create flexibility regarding potential opponents). Table \ref{tab:grouprecommendationtennispractices} depicts an example training plan recommendation (training partner assignment) for two weeks. Such recommendations can be determined, for example, with constraint-based recommendation where the expertise difference between training partners can be the basis for a corresponding optimization function (see Formula \ref{tab:optimizetennispartnerassignment}).

 \end{example}

 \begin{equation} \label{tab:optimizetennispartnerassignment}
     MIN \leftarrow \Sigma_{(pl_i,pl_j) \in w_1} |itn(pl_i)-itn(pl_j)| +  \Sigma_{(pl_i,pl_j) \in w_2} |itn(pl_i)-itn(pl_j)|
 \end{equation}

\begin{table}[ht]
\centering \caption{A simplified example of training partner (player $pl_i$) assignment for weekly trainings. In this context, \emph{itn} represents the expertise level (the lower the better).}
\begin{tabular}{|c|c|c|c|c|c|c|c|c|c|c|c|c|c|c|c|c|c|c|c|c|c|c|c|c|c|} 
\hline
  \multicolumn{2}{|c|}{team} & \multicolumn{6}{c|}{week$_1$ ($w_1$)}  & \multicolumn{6}{c|}{week$_2$  ($w_2$)}      \tabularnewline \hline 
player & itn  & $pl_1$  & $pl_2$   & $pl_3$ & $pl_4$  & $pl_5$ & $pl_6$ & $pl_1$  & $pl_2$  & $pl_3$  & $pl_4$ & $pl_5$  & $pl_6$\tabularnewline \hline \hline
 pl$_1$ & 5.8 & -       & \bf x        & -      & -       & -      & -      & -       & -       & -       & \bf x      & -       & -  \tabularnewline \hline
 pl$_2$ & 6.2 & \bf x       & -        & -      & -       & -      & -      & -       & -       & \bf x       & -      & -       & - \tabularnewline \hline
 pl$_3$ & 6.4 & -       & -        & -      & -       & -      & \bf x      & -       & \bf x       & -       & -      & -       & - \tabularnewline \hline
 pl$_4$ & 6.5 & -       & -        & -      & -       & \bf x      & -      & \bf x       & -       & -       & -      & -       & - \tabularnewline \hline
 pl$_5$ & 6.7 & -       & -        & -      & \bf x       & -      & -      & -       & -       & -       & -      & -       & \bf x \tabularnewline \hline
 pl$_6$ & 6.9 & -       & -        & \bf x      & -       & -      & -      & -       & -       & -       & -      & \bf x       & - \tabularnewline \hline
\end{tabular} 
\label{tab:grouprecommendationtennispractices} 
\end{table}

\subsubsection{Focused Trainings}

Focused trainings follow the idea of improving specific techniques relevant in a specific sports discipline. Kashino \cite{Kashino2018Brain} emphasizes the importance of trainings that help to improve cognitive functions for better grasping situations, taking better instantaneous decisions, and also remaining in a kind of optimal mental state even in extremely challenging situations. In this line of research, Roanes-Lozano et al. \cite{RoanesLozanoetal2020} introduce a constraint-based recommendation approach focusing on the improvement of a tennis serving technique where a users current state of practice is found out on the basis of question/answer dialog. With the collected information, corresponding measures are determined on the basis of constraints matching user answers with potential service improvement trainings.  Finally, Portaz et al. \cite{Portazetal2023} introduce an approach to the CF-based recommendation of psychomotor training practices in various contexts such as physical rehabilitation, keeping active when getting older, and different types of basic performance improvements goals also including professionals seeking to increase their excellence. 

\begin{example}[Recommendation of Tennis Practices] Training sessions in tennis are often related to practicing in the context of specific game situations, for example, a session could start with forehand cross playing and -- as soon as one opponent (training partner) plays a short ball -- an attack of the other partner follows with the goal to successfully win the rally.  In Table \ref{tab:tennistrainingsessions} (repository of training sessions), such a focused training episode is represented by the training session \emph{forehand cross \& attack}. The table entries represent weights regarding the capability of a specific training session to improve specific skills, for example, training session $2$ can help to significantly improve the quality of a players \emph{forehand}. 
\end{example}

\begin{table}[ht]
\centering \caption{Tennis training sessions as items to be recommended. For example, training session $2$ contributes to forehand improvements but not necessarily to significant volley improvements.}
\begin{tabular}{|c|c|c|c|c|c|c|c|c|c|c|c|c|c|c|c|c|c|c|} 
\hline
  id   & training session           &   forehand   &    backhand    &    volley     &    serve \tabularnewline \hline
  $1$  &  serve \& volley           &    0.1       &       0.1      &      0.2      &    0.6 \tabularnewline \hline
  \bf  $2$  &  \bf  forehand cross \& attack  &    0.7       &       0.25     &      0.05     &    0.0 \tabularnewline \hline
  $3$  &  backhand cross \& attack  &    0.25      &       0.7      &      0.05     &    0.0 \tabularnewline \hline
  ..   &  ..                        &    ..        &       ..       &      ..       &    .. \tabularnewline \hline
\end{tabular} 
\label{tab:tennistrainingsessions} 
\end{table}

Depending on the preferences of a player regarding the improvement of his/her skills, different training sessions can be recommended, for example, by using a utility-based recommendation approach. If a player is primarily interested in improving his/her forehand, this can be represented with the weight vector (forhand: 1.0, backhand:0.0, volley:0.0, serve:0.0). If we apply these weights to the  training sessions, \emph{forehand cross \& attack} would  have the highest evaluation which is $0.7$ ($1.0 \times 0.7 + 0.0 \times 0.25 + 0.0 \times 0.05 \times 0.0 \times 0.0 = 0.7$). Such a utility-based approach is often used in the context of constraint-based recommendation, where recommendation candidates (consistent with a given set of constraints) are ranked according to their utility for the user \cite{FelfernigBurke2008constraintbased}. This scenario could easily be extended to scenarios where individual team members have to be taken into account on the basis of group-based recommendation \cite{Felfernig2018_1}. In tennis, this would mean to take into account the training preferences (needs) of two training partners. In our example, we only used one set of preferences, i.e., no collection of  individual preferences.

\subsubsection{Sports Equipment}

In the majority of sports disciplines, the used equipment has a significant impact on various factors such as performance and risk. For example, running shoes have an impact on the marathon running performance and high-quality climbing equipment is also an important factor to assure an athletes safety. Depending on the personal preferences (e.g., light-weight and flexible running shoes vs. solid ones) and goals (e.g., running 10 kilometres vs. running a marathon), different sports items must be recommended. Although the application of collaborative filtering (for the exploration of new equipment) and content-based filtering (for the recommendation of new but very similar equipment) is reasonable, constraint-based  recommendation approaches can better help to take into account goal-specific restrictions \cite{Felfernig2006,LoChangSheuJung2014} which cannot be implemented on the basis of content-based and collaborative approaches. For example, a heavy-weight tennis racket should not be recommended to beginners and players with a focus on playing doubles. Table \ref{tab:KnowledgeBasedRecommendationCandidateItemsRackets} depicts an assortment of tennis rackets which can be recommended on the basis of a constraint-based recommendation approach.

\begin{example}[Recommendation of Tennis Rackets] Tennis material recommendation could propose tennis rackets better supporting a players strengths. Recommending such items often comes along with a ramp-up problem since only limited data (if at all) is available for determining recommendations. A related solution is to follow a knowledge-based approach where recommendations are determined, for example, with  rules and utility scores. If a user knows his/her preferred playing style, knowledge-based recommenders can help to identify relevant items (Table \ref{tab:KnowledgeBasedRecommendationCandidateItemsRackets} includes a set of tennis rackets with the basic properties, for example, \emph{balance point}, \emph{weight}, and \emph{price}). 
\end{example}

Allowing extreme top-spin performances requires a "top-heavy" racket. With such rules, constraint-based recommenders can narrow down the search space. The remaining candidates can be ranked with regard to a predefined set of interest dimensions. Let us assume that a recommender has identified three candidate items (see Table \ref{tab:KnowledgeBasedRecommendationCandidateItemsRackets}).  In this example, a constraint could specify that if a user is interested in high top-spin performance, only rackets with at least a balance factor of $325$ should be recommended. This can be specified in terms of a constraint such as $playinggoal=topspin \rightarrow balance \geq 325$.

\paragraph{} Following the scheme of Table \ref{tab:utilitiesoftennisrackets}, the items of Table \ref{tab:KnowledgeBasedRecommendationCandidateItemsRackets} can then be ranked according to the interest dimensions \emph{quality} (the better the top-spin performance, the higher the quality) and \emph{economy} (the lower the price, the higher the economy). Since company $B$ produces higher-quality rackets (with a higher price), it has a lower economy-related evaluation. The higher the racket balance value, the higher the top-spin potential and the higher the contribution to the interest dimension \emph{quality}.  

\begin{table}[ht]
\centering \caption{An example collection of tennis rackets.}
\begin{tabular}{|c|c|c|c|c|c|} 
\hline
  item (tennis racket) & company & balance & weight & price   \tabularnewline  \hline \hline
  \bf racket$_1$           & \bf A       & \bf 325     & \bf 285    & \bf 100     \tabularnewline  \hline
  racket$_2$           & B       & 325     & 300    & 100     \tabularnewline  \hline
  racket$_3$           & B       & 340     & 300    & 200     \tabularnewline  \hline
\end{tabular} 
\label{tab:KnowledgeBasedRecommendationCandidateItemsRackets} 
\end{table}

\paragraph{} In our scenario, economy($racket_1$) = 10+10+10+10 = 40, economy($racket_2$) = 34, and economy($racket_3$) = 27 whereas quality($racket_1$) = 5+7+7+7 = 26, quality($racket_2$) = 34, and quality($racket_3$) = 40. Assuming an equal importance of the interest dimensions \emph{economy} and \emph{quality}, this would result in an overall utility (i.e., economy + quality evaluation of an item) of $racket_1=40+26=76$, $racket_2=34+27=61$, and $racket_3=40+26=67$, i.e., $racket_1$ should be recommended first. For sure, variations in the importance of interest dimensions would lead to different recommendations.

\begin{table}[ht]
\centering \caption{A simple utility-based evaluation scheme for tennis racket recommendation.}
\begin{tabular}{|c|c|c|c|c|c|c|c|c|c|c|c|c|c|c|c|c|c|c|} 
\hline
  interest dimension & \multicolumn{2}{c|}{company}  & \multicolumn{2}{c|}{balance}  & \multicolumn{2}{c|}{weight}  & \multicolumn{2}{c|}{price}    \tabularnewline \cline{2-9} 
           & A  & B   & 325 & 340   & 285  & 300  & 100  & 200 \tabularnewline \hline \hline
  economy  & 10 & 5   & 10  & 8     & 10   & 9    & 10  & 5 \tabularnewline \hline
  quality  & 5  & 10  & 7   & 10    & 7    & 10   & 7  & 10 \tabularnewline \hline
\end{tabular} 
\label{tab:utilitiesoftennisrackets} 
\end{table}

\subsubsection{Videos}

A major strength of sports video recommendations is their capability of demonstrating in a user-adaptive fashion specific practices in such a way that the personal performance can be improved. For example, a slow-motion study of the backhand of an ATP tennis player or a detailed demonstration of the motion sequence of workouts \cite{Ezin2018FitnessTF,Odden2017} can help to more easily "reproduce" a specific motion pattern. In this context, collaborative filtering can help to exploit serendipity effects, for example, in the context of watching videos for improving ones own tennis service, training sessions regarding smash practices could be recommended, since both moves have the same underlying motion patterns. In contrast, content-based filtering will help to find videos more focusing on the same topic a user was interested in previous recommendation sessions. Video recommendations are regarded as a valuable alternative to traditional onsite demonstration training since individual demonstrations can be repeated on-demand \cite{SunTangSportsDanceVideo2022}.

\subsubsection{Destinations \& Routes}

Recommendations of destinations and routes can be found in various sports-related application contexts.  Avesani et al. \cite{Avesanietal2005} introduce a CF-based approach to the recommendation of \emph{ski routes}. Such scenarios have to deal with the specific challenge of abrupt changes in weather conditions -- in this context, user-specific trust measures have been integrated (e.g., in terms of other users trustworthiness evaluations)  with the goal to increase the credibility of recommendations. Ivanova et al. \cite{Ivanovaetal2022,IvanovaIntroducingContext2023,IvanovaWald2023,LoeppZiegler2020} introduce different CBF recommendation scenarios in the context of outdoor activities such as hiking, running, and climbing. Specifically, the authors focus on the recommendation of climbing cracks and different routes taking into account a user context specified, for example, in terms of time of the year or persons who participate (e.g., family vs. a team of professional climbers). A content-based approach to the recommendation of walking routes is presented in Li et al. \cite{Lietal2021} where user preferences and walking routes are matched by taking into account sentiments, i.e., the overall sentiment (more specifically, an emotional score) associated with items depending on a social network based tweet analysis. Further content-based recommendation approaches in the context of route recommendation are presented in Takama et al. \cite{Takamaetal2015} (walking routes) and Teslyuk et al. \cite{Teslyuketal2019} (bike routes).  An approach to the content-based recommendation of available locations and courts for user-preferred activities (e.g., tennis and futsal) is presented in Othman et al. \cite{Othmanetal2023}. Another related example is the recommendation of skiing resorts where basic user training requirements (e.g., skiing expertise and intended type of skiing) are matched with corresponding skiing resorts \cite{Serdouketal2021}. Following the concepts of group recommender systems, McCarthy et al. \cite{McCarthyetalSkiGroupRec2006} introduce an approach to the critiquing-based recommendation of skiing resorts where a group of friends wants to take a decision regarding the skiing resort to be visited. Team members can define their preferences with a critiquing-based user interface. A group recommender then tries to aggregate those preferences in such a way that all of the group members are satisfied with the recommendation. In a different way, Wirz et al. \cite{Wirzetal2011} introduce an approach to interpret individual paragliding behaviors as collective (group) behavior which is then used for the recommendation of thermal spots supporting excellent paragliding experiences.

\subsubsection{E-Sports Related Aspects}
Recommendations in e-sports, i.e., online games that are played in a highly competitive environment, propose relevant games (e.g., the next game of relevance for a player), game settings (e.g., game maps and opponents in a specific competitive scenario), and in-game content choices to players (e.g., weapons and additional equipment) and also partners and opponents \cite{Wuetal2017}. In this context, various criteria are taken into account ranging from an appropriate game difficulty to an attractive environment (e.g., in terms of suitable map recommendations). Proposing the most suitable game scenario (e.g., a map) and a corresponding game mode is also important for a good onboarding experience and to retain players \cite{Wuetal2017}.

\subsection{Tactical \& Strategic Planning} \label{subsec:tacticalplanning}

With \emph{tactical \& strategic planning} planning we refer to two different planning levels which are both crucial for sustainable sports-related goal achievement \cite{Lewis2004}. They differ in terms of the basic goal (e.g., winning a soccer game using an offensive or a defensive tactic vs. winning a whole soccer championship), time frame (e.g., preparation for a specific game vs. planning a team composition for the whole season), and engaged persons (e.g., team coach and players discuss the lineup for the next match vs. management and coach discuss the team playing direction for the next season). Related recommendations can be provided directly to an athlete or used as a basis for in e-coaching environments mediating between athletes and physical coaches \cite{BorattoetalPredictingWorkoutQuality2018,BorattoetalECoachingEcoSystem2017}. In the following, we focus on tactics recommendation.

\subsubsection{Tactics in Endurance Sports}

In the context of marathon running, it is important to select optimal pacings for different phases of an upcoming marathon run. This is comparable to  decisions in other endurance sports such as cross-country skiing, cycling, and swimming.  In  marathon runs, case-based recommender systems can help to identify relevant race tactics (in terms of pacings) by identifying athletes with similar performances   \cite{Berndsenetal2019,SmythCunningham2017,Smyth2022}.  In the recommendation scenario depicted in Table \ref{tab:casebasedmarathonpacerecommendation}, runner$_1$ is the marathon runner who successfully completed a marathon with a running time similar to the envisioned running time of the "current" runner and also with similar initial pacings in kilometres 1--10. The pacings of runner$_1$ on kilometres $11-42$ can be used as recommendations for the current runner.

\begin{table}[ht]
\centering \caption{Recommendation scenario of marathon-specific pacings (per kilometre) using case-based recommendation (CBR). In this context, $k_{\alpha-\beta}$ denotes a running segement starting with kilometre $\alpha$ and ending with kilometre $\beta$.}
\begin{tabular}{|c|c|c|c|c|c|c|c|c|c|c|c|c|c|c|c|c|c|c|} 
\hline
              & $k_{1-5}$  & $k_{6-10}$   & $k_{11-15}$ & $k_{15-20}$  & ... & $k_{36-42}$ & time (hours) \tabularnewline \hline \hline
  runner$_1$  & 6.5 & 6.5   & \bf 6.75  & \bf 6.75     & ...   & \bf 6.5  &  \bf 4.5   \tabularnewline \hline
  runner$_2$  & 5.5  & 5.0  & 5.0   & 5.0    & ...    & 4.5   & 3.5  \tabularnewline \hline \hline
  current  &   6.75  & 6.5  &  ?  & ?    & ?    & ? & 4.5   \tabularnewline \hline
\end{tabular} 
\label{tab:casebasedmarathonpacerecommendation} 
\end{table}

\subsubsection{Tactics Against Opponents}

A  way to select specific tactics is to watch  videos and understand in which way opponents "implement" specific tactics (e.g., offensive vs. defensive). In this context, it is important to analyse videos with regard to different behavioral patterns of opponents and thus to better support game tactics preparation for a team \cite{WuetalVideoActionRecognition2022}. Tsai et al. \cite{TsaietalTacticRecognition2017}  address to problem of tactic recognition in basketball videos. An important task in this context is to identify individual players, understand their temporal variations, and aggregate the identified information into a classification of the underlying tactic (e.g., offensive vs. defensive). Based on information about tactics in different video segments, a content-based recommender system can search for similar game situations with the same  or other opponents.

From the viewpoint of an athlete or a team, it is highly relevant to be able to estimate the appropriateness of different tactics in an upcoming competition. Related work is reported, for example, in the context of developing tactics in soccer scenarios (e.g., to identify a  tactics of the next game \cite{AbreuetalCF2014,Abreuetal2014,Mengetal2020}) and table tennis (e.g., tactics recommendation for the next table tennis match in an international competition \cite{Sunetal2020TableTennisTactics}). In such scenarios, collaborative filtering can help to identify teams which were similar successful in the past and also followed similar game tactics. With this information, CF can  identify similar teams/athletes (the nearest neighbors) and to choose a promising tactic for an upcoming competition. This information can also be used to predict the outcome of specific matches if a team has already decided on a specific tactics for the upcoming match.  

Chang and Qiu \cite{ChangYu2022TennisTactics} introduce an approach to the inference of decision trees from tennis player strengths and weaknesses, for example, in terms of  hitting techniques (forehand, backhand, volley, serve, slide, topspin, and smash), physical condition, and habitual scoring positions. Based on a corresponding dataset, the authors show how to derive a decision tree which can be regarded as a constraint set representing tactics in specific gaming situations.  Examples of basic rules in the context of tennis matches are the following: \emph{if opponent = X and personal 1st service rate $>$ 70\%, focus on intensive net attacks} and \emph{if opponent = defender, focus on high-speed ground strokes}.

\begin{example}[Recommendation of Team Tactics]
A simplified example of the application of CF in team tactics recommendation is depicted in Table \ref{tab:grouprecommendationteamtactics}. This scenario is based on the  assumption that there are two possible game tactics which are (1) defensive (d) and (2) offensive (o). 
\end{example}

\begin{table}[ht]
\centering \caption{A simplified example of a collaborative filtering based team tactics recommendation (d=defensive whereas o=offensive tactics).}
\begin{tabular}{|c|c|c|c|c|c|c|c|c|c|c|c|c|c|c|c|c|c|c|c|c|c|c|c|c|c|} 
\hline
             & $team_1$         & $team_2$          & $team_3$    & ... &$team_n$    \tabularnewline \hline 
  $team_1$   &   -              &  3.0 (o)          &  0.0   (o)  & ... &    -       \tabularnewline \hline 
  $team_2$   &   3.0   (d)      &  -                &  0.0 (d)    & ... &  0.0 (d)   \tabularnewline \hline 
   $\bf team_3$   &   5.0   (d)      &  5.0 (o)          &  -          & ... &  -         \tabularnewline \hline 
  ...        &     ...          &     ...           &    ...      & ... &  ...       \tabularnewline \hline 
  $team_{current}$   &   -              &  5.0 (o)          &   ?         & ... &  -         \tabularnewline \hline 
\end{tabular} 
\label{tab:grouprecommendationteamtactics} 
\end{table}

Let us assume that the upcoming soccer game for team$_{current}$ will we be against team$_1$. The nearest neighbor of team$_{current}$ in terms of used strategy and degree of success ($loss=0.0$ .. $win=5.0$) is team$_3$ which also won against team$_2$ with the same (or a similar) tactics. Since the nearest neighbor won against team$_1$ following a defensive tactics (d), such a tactics could also be recommended to team$_{current}$. In real-world settings, the availability on specific players on specific positions and more detailed knowledge about the strengths and weaknesses of opponents have to be taken into account to infer reasonable tactics.

\begin{example}[Recommendation of Soccer Tactics]
Tactics selection can be supported with group recommendations (see Table \ref{tab:grouprecommendationsoccerstrategy}). Depending on the  available players of a soccer team, different strategies may be more or less appropriate. Strategies such as offensive, defensive, pressing, one-on-one defense, and combinations thereof can be selected by the coaching staff. In Table \ref{tab:grouprecommendationsoccerstrategy}, the coaching staff consisting of coaches coach$_1$ .. coach$_3$ is in charge of deciding about the playing tactics against the next opponent -- coach$_3$ is assumed to be a virtual agent capable of estimating the relevance of different game tactics -- for a related overview, we refer to Beal et al. \cite{Bealetal2019}.
\end{example}   

\begin{table}[ht]
\centering \caption{Simple example of soccer tactics recommendation with group recommendation. Each coach evaluates the available tactics alternatives and a scale $1..5$ (the higher the better).}
\begin{tabular}{|c|c|c|c|c|c|c|c|c|c|c|c|c|c|c|c|c|c|c|c|c|c|c|c|c|c|} 
\hline
            & \multicolumn{4}{c|}{game tactics alternatives} \tabularnewline \cline{2-5} 
     staff  & offensive & offensive $\times$ pressing & defensive & defensive $\times$ one-to-one \tabularnewline \hline \hline
  coach$_1$ &   1       &         0                   &    4      &     5                         \tabularnewline \hline
  coach$_2$ &   0       &         0                   &    4      &     2                         \tabularnewline \hline
  coach$_3$ (virtual) &   2       &         0                   &    4      &     4                         \tabularnewline \hline \hline
  AVG       &   1       &         0                   &    \bf 4      &     3.6                       \tabularnewline \hline
\end{tabular} 
\label{tab:grouprecommendationsoccerstrategy} 
\end{table}

In the scenario depicted in Table \ref{tab:grouprecommendationsoccerstrategy}, a group recommender is just used to select the preferences of the individual coaches and -- on the basis of the collected feedback -- recommends the globally preferred tactics which can then be used for further discussions. In order to aggregate the individual preferences of the coaches, the \emph{average} (AVG) aggregation function is used.\footnote{For a detailed discussion of aggregation strategies, we refer to Felfernig et al. \cite{Felfernig2018_1}.}  One adaptation to this basic aggregation function would be a kind of weighted average giving the head coach a "louder voice". When taking a look at the feedback of the individual coaches in Table \ref{tab:grouprecommendationsoccerstrategy}, a basic \emph{defensive tactics} could be regarded as the preferred one, however, a recommender should also point out diverging evaluations which should be further discussed  by the coaches, for example, there seems to be dissent regarding the \emph{defensive} $\times$ \emph{one-to-one} approach and it might still be the case that this tactic is finally selected as the "winner" although having initially received a lower evaluation.

\subsubsection{E-Coaching}

The idea of e-coaching is to provide coaching support on the basis of a communication platform where performance data (e.g., of training practices) of sports people is stored and coaches have access and thus can estimate the progress made. In such scenarios, it is important to support coaches in terms of (1) choosing sportsmen for whom coaching interventions are crucial to enable improved performance and  (2) proposing relevant measures, for example, in terms of adapted training practices or simply some motivating messages \cite{BorattoetalPredictingWorkoutQuality2018}. Sportsmen coaching sessions need to be prioritized and coaching support needs to be provided timely. In this context, fairness plays a crucial role, i.e., sportsmen should have somehow an equal opportunity of being supported by a coach \cite{Borattoetal2022eCoaching,BorattoetalECoachingEcoSystem2017}. A simplified example of how to take into account fairness in such scenarios is depicted in Table \ref{tab:examplefairnessesports} where individual sportsmen are evaluated with regard to the aspects of coaching support time consumed up to now, criticality of coaching support, and potential for improvement.

\begin{table}[ht]
\centering \caption{A simple example of a fairness-aware coaching session recommendation setting where \emph{time} represents the already consumed coaching time (in minutes). In this context, $rank=1$ indicates the sportsman to be supported next.}
\begin{tabular}{|c|c|c|c|c|c|c|c|c|c|c|c|c|c|c|c|c|c|c|c|c|c|c|c|c|c|} 
\hline
     sportsman  & time   & criticality (1..10)    & potential  (1..10) & relevance ($s_i$) & rank \tabularnewline \hline
     $s_1$      &   50   &  5                     & 5                  & 0.5               & 3    \tabularnewline \hline
     $s_2$      &   10   &  8                     & 9                  & 7.2               & \bf 1    \tabularnewline \hline
     $s_3$      &   30   &  6                     & 7                  & 1.4               & 2    \tabularnewline \hline
\end{tabular} 
\label{tab:examplefairnessesports} 
\end{table}

On the basis of the entries in Table \ref{tab:examplefairnessesports}, we are able to derive a corresponding fairness-aware evaluation of the relevance of coaching support for individual sportsmen (see Formula \ref{eq:ecoaching}). The relevance of coaching support increases with an increasing potential of the sportsman and an increasing criticality of coaching support. At the same time, the relevance decreases with an increasing coaching time amount consumed by a sportsman. This is a simple way of taking into account fairness when determining a relevance ranking for individual coaching sessions.
   
\begin{equation}\label{eq:ecoaching}
    relevance(s)=\frac{potential(s) \times criticality(s)}{time(s)}
\end{equation}

\subsection{Recommending Individuals \& Teams}  \label{subsec:teamconfiguration}

Recommending individual athletes or teams is a complex task including different aspects such as (1) the recommendation of talents with the skills required to be a successful athlete taking into account potential future injuries (inferred from issues in the past), (2) the configuration of teams (in the short and long run) which are able to support tactics and strategies envisioned by coaches, (3) the achievement of team dynamics in terms of athletes who can cooperate with each from the personality point of view, (4) optimized long-term operational planning to achieve maximum output without activating injuries (5) and assuring marketability of athletes, for example,  taking into account an athletes popularity in the fan base.

\subsubsection{Individuals}

Information about the individual strengths and weaknesses of players can help to recommend individual players and whole teams \cite{Chenetal2018HeroRecommendation,Patil2020}. Klancar et al. \cite{Klancaretal2019} show how to apply collaborative filtering (CF) concepts in e-sports scenarios. After a knowledge-based pre-filtering, CF determines recommendations for players that should be part of the virtual team. Tu et al. \cite{Tuetal2017} present a CF based approach that helps to identify partners with a fit with regard to an activity (e.g., a tennis training session). The idea behind is that basic CF can help to identify partners with similar interests which can be a natural approach in terms of finding partners for a new activity. With the goal of game character selection, B\l{}aszczyk and Szajerman \cite{Blaszczyketal2023} present a CF approach developed for the Multiplayer Online Battle Arena (MOBA) e-sports game League of Legends (LoL) which has over 160 champions, i.e., playable characters, to select from at the beginning of every match. As champions have diverse abilities and characteristics, recommendation features are needed to make the identification of preferred champions less effortful. In this context, B\l{}aszczyk and Szajerman \cite{Blaszczyketal2023} introduce a CF-based recommender system which recommends champions on the basis of champions that have been selected previously by the same user.

Video analysis plays a major role when trying to identify talents. Using such information, a preference profile defined by a team management can be matched with the identified video contents and video analysis processes can be completed more efficiently \cite{WuetalVideoActionRecognition2022}. As an orthogonal means compared to video analysis, IoT infrastructures \cite{FelfernigetalIOT2019} can help to record an athletes activities and -- on the basis of this information -- to infer a  behavior model \cite{IvanovaBehavior2021}. Such a model can then be used to match an athletes capabilities (e.g., acceleration signals of climbers) with requirements of sports management. Furthermore, trainings can be better personalized and thus help to achieve a sustainable development of a talents capabilities.

For young athletes, it is extremely important to have a clear picture of their strengths and weaknesses and a corresponding support when choosing a specific sports discipline. Such a decision scenario can be supported, for example, with constraint-based recommenders which match (on the basis of constraints) information from motoric skills tests and morphologic characteristics functional tests with a corresponding reasonable type of sports \cite{Papicetal2009TalentIdentification}. A simple example of such a constraint is the following: if a person is interested in \emph{swimming}, this implies that the persons \emph{height} is \emph{tall} and the \emph{BMI} is \emph{semi-low} \cite{Papicetal2009TalentIdentification}.

\begin{example}[Recommendation of Players] CF can predict the  appropriateness of a candidate player for a team (see Table \ref{tab:playerCFrecommendation}) \cite{Klancaretal2019}. 
\end{example}

\begin{table}[ht]
\centering \caption{Simple example of CF-based player recommendation: in this context $pl_\alpha$ represent different player types which were engaged in different (successfully completed) games, for example, in game$_1$, among others, player types \{$pl_1$,$pl_3$,$pl_n$\} were engaged.}
\begin{tabular}{|c|c|c|c|c|c|c|c|c|c|c|c|c|c|c|c|c|c|c|c|c|c|c|c|c|c|} 
\hline
     teams      & $pl_1$ & $pl_2$   &  $pl_3$ &    ..      & $pl_n$ \tabularnewline \hline \hline
  game$_1$      &   x       &   -   &    x    &    ..      &     \bf x  \tabularnewline \hline
  game$_2$      &   -       &   x   &         &    ..      &     -  \tabularnewline \hline
  game$_3$      &   -       &   -   &         &    ..      &     -  \tabularnewline \hline \hline
  next game     &   x       &  ?    &    x     &    ..      &     ?  \tabularnewline \hline
\end{tabular} 
\label{tab:playerCFrecommendation} 
\end{table}

If we assume that $pl_\alpha$ in Table \ref{tab:playerCFrecommendation} represent player types, a CF-based recommendation service can be of great value in the identification of players helpful for the next game. In our example, player type $pl_n$ could be regarded as relevant for the next game since it also contributed to a successful completion of game$_1$. 

\subsubsection{Teams}

Team recommendation is about proposing team member configurations that help to succeed in an upcoming competition. Dependencies between individual team members are often represented in terms of constraints, for example, player A and B have a longterm experience in playing together, i.e., this synergy effect should be exploited in terms of integrating both players. Using a constraint-based recommendation, Rajesh et al. \cite{Rajeshetal2022} introduce a team recommendation approach in the FPL fantasy (football) premier league. In such settings, constraints refer to aspects such as the number of required players, the overall available budget, "to-be-filled" positions, and compatibilities between individual players \cite{LandersDuperouzel2019,Thavamunietal2023TeamConfiguration}. In the following example, we show how tennis player selection can be "implemented" on the basis of a constraint-based recommendation approach.

\begin{example}[Recommendation of Tennis Teams] In the previous example, we sketched an approach to determine the appropriateness of individual player types for a team. Going one step further, we can apply knowledge-based recommendation to configure a whole team. If we assume the availability of data describing previously played games, such a dataset could include  relevant information regarding the compatibility of individual players, the appropriateness of players for individual positions, and the current performance level of individual players. On the basis of such a dataset, we are able to define a basic team configuration task as follows -- in this simplified example, we only take into account the aspects of player compatibilities and current individual performance level (see Table \ref{tab:soccerplayercompatibilities} for an example of a tennis team configuration). 
\end{example}

\begin{table}[ht]
\centering \caption{Simple example of the definition of compatibilities between tennis players.}
\begin{tabular}{|c|c|c|c|c|c|c|c|c|c|c|c|c|c|c|c|c|c|c|c|c|c|c|c|c|c|} 
\hline
     teams  & level (l) & $pl_1$ & $pl_2$   &  $pl_3$    \tabularnewline \hline \hline
  pl$_1$    &  4    &   5       &   3.0   &    4.0   \tabularnewline \hline
  pl$_2$    &  5    &   3.0     &   5     &    4.0   \tabularnewline \hline
  pl$_3$    &  4    &   4.0     &   4.0   &    5     \tabularnewline \hline 
\end{tabular} 
\label{tab:soccerplayercompatibilities} 
\end{table}

Based on data as defined in Table \ref{tab:soccerplayercompatibilities}, a tennis team consisting of two persons can be identified on the basis of the following utility function (see Formulae \ref{eq:argmaxtennisteams}--\ref{eq:utilitytennisteam}) that determines the team with the overall maximum utility which is the team \{$pl_2,pl_3$\} (utility($pl_1,pl_2$)=3.0 $\times$ (4+5)=27, utility($pl_1,pl_3$)=4.0 $\times$ (4+4)=32, and utility($pl_2,pl_3$)=4.0 $\times$ (5+4)=36).

\begin{equation}\label{eq:argmaxtennisteams}
   \arg\max_{(pl_i,pl_j)}  [utility(pl_i,pl_j)] ~(i \neq j)
\end{equation}

\begin{equation}\label{eq:utilitytennisteam} 
   utility(pl_i,pl_j) = compatibility(pl_i, pl_j) \times (level(pl_i) + level(pl_j))
\end{equation}

Following the concepts of \emph{collective intelligence}, such basic approaches to identifying teams can be extended with regard to various dimensions \cite{Chmait2017}. For example, team recommendations could be determined not just for a single competition but also for a series of competitions which makes it important to also take into account the physical load of individual players (e.g., in the context of football world championships). Furthermore, team configuration can also take into account properties of opponents, for example, in the context of our tennis scenario, the playing level $l$ is just a single number representing the qualities of a player. However, this can be arranged in a more fine-grained fashion by analyzing in more detail how well individual players can react on the strengths of opponents. 

 A scenario to apply case-based recommendation in the context of team recommendation is to support substitute player selection, where alternative substitutes can be regarded as cases and substitutes most similar to the current player are regarded as primary recommendation candidates \cite{YIlmazetal2022}. In soccer, examples of related attributes for estimating the similarity between individual players are accuracy in passing, successful dribblings, sprint speed, and acceleration, in cricket, examples of such attributes are batting average and batting strike rate \cite{Jayanthetal2018}. A generalization of the substitute recommendation scenario is the recommendation of team completions, for example, on the basis of a defined set of regular players, a case-based recommender could recommend team completions from similar successful team configurations in the past.

\subsection{Handling Injury \& Illness} \label{subsec:injuryhandling}

Recommender systems can provide advisory services regarding the avoidance of  and the recovery from injuries \cite{Bealetal2019,Eetveldeetal2021,ZhangRiskAssessment2022}. By taking into account an athletes physical condition and also historical data, recommender systems can  propose injury avoidance measures (e.g., in terms of adjustments in training programs, loads, and training-free periods) and measures regarding rehabilitation (e.g., in terms of identifying appropriate medical support, recovery exercises, and needed equipment such as bandages). 

\subsubsection{Measures for Injury/Illness Avoidance}

Physical activity helps the immune system to be more resistant regarding respiratory diseases. Bhimavarapu et al. \cite{Bhimavarapuetal2021} introduce a collaborative system focusing on the recommendation of potentially relevant physical activities based on a persons health status and further information including eating and exercise habits. With the goal of preventing such kinds of injuries, recommender systems can propose articles on potential consequences of using too heavy-weight tennis rackets as well as overstressing joints due to too frequent training sessions. As reported in Emrich et al. \cite{Emrichetal2014}, content-based recommendation is also applied to support the recommendation of "health-aware" running or biking routes by matching  a persons preferences (e.g., in termos of locations) but also the physical condition with route properties (e.g., "hilliness").  Similar to the recommendation of  training sessions taking into account specific constraints such as  current physical condition \cite{SantosGagoetal2019RecommenderSportswomenConstraintBasedMenst}, constraint-based recommendation can also be used to avoid injuries caused, for example, by too intensive training sessions \cite{Matosetal2020}. Finally, Yang et al. \cite{Yangetal2018} present an approach to depression prediction combined with a constraint-based recommendation of emotional improvement suggestions guiding a users behavior. 

\begin{example}[Prediction of Injuries]
Given a set of observations and training data, the duration of recovery and the probability of  injuries could be predicted (see the example in Table \ref{tab:predictinginjuries}) \cite{Feelyetal2021}. In this example, we are able to predict potential issues of the current runner. 
\end{example}

\begin{table}[ht]
\centering \caption{Simple example of applying case-based recommendation in recovery prediction based on weekly observations of individual runners.}
\begin{tabular}{|c|c|c|c|c|c|c|c|c|c|c|c|c|c|c|c|c|c|c|c|c|c|c|c|c|c|} 
\hline
runner         & age    &   practising hours   & distance  &   injury      \tabularnewline \hline \hline
runner$_1$     &   30   &     6                &   54      &      n        \tabularnewline \hline
runner$_2$     &   22   &     6                &   71      &      y        \tabularnewline \hline
runner$_3$     &   25   &     1                &   2       &      \bf y        \tabularnewline \hline \hline
current        &   35   &     1                &   1       &      ?        \tabularnewline \hline
\end{tabular} 
\label{tab:predictinginjuries} 
\end{table}

Let us assume that runner$_3$ is the nearest neighbor of \emph{current} which could lead us the the conclusion that runner \emph{current} is somehow in danger of an injury, for example, due to the fact that nearly no training run took place in the last week. Having such indications helps to better to generate corresponding explanations making the current situation transparent. In contrast to a low training level, also a too intensive training could lead to injuries -- for details see \cite{Feelyetal2021,Kluitenberg2015}. Arciniega-Rocha et al. \cite{Arciniegaetal2023} present an approach to the application of  case-based recommendation helping to avoid injuries in the workouts of amateur athletes. Since even experienced trainers are unable to detect each suboptimal movement "manually", wearable devices are used as a basis for data collection of experienced athletes. By finding similar cases, i.e., experienced athletes, and systematically analyzing differences compared to the "person under observation", specific recommendations can be determined based on the identified differences in the collected workout data.

\subsubsection{Recovery Actions}

During rehabilitation, exercise games (so-called "exergames") help to encourage patients without feeling bored. For such scenarios,  Ismail et al. \cite{Ismailetal2021} propose a collaborative filtering based recommendation system that proposes exercise settings which are most suitable to optimally improve a patients rehabilitation success. With this approach, for the current patient nearest neighbors with similar injuries and rehabilitation phases are identified and the most promising exercises performed by those patients are recommended to the current patient. Before even entering a rehabilitation process, patients are in the need of identifying the best-suited doctors for supporting such a process -- Shambour et al. \cite{Shambourteal2023}  propose a hybrid recommendation approach where doctors are also recommended on the basis of the selected doctors of nearest neighbors. 

Content-based recommendation can help to find important information, for example, in terms of books, articles, and videos, to support injury and illness avoidance and recovery \cite{RanaetalContentBasedHealthRecommenderSystems2020}. For example, so-called tennis-elbows are a wide-spread issue in tennis sports -- being confronted with such issues, users would be interested in ways to recover and get back to tennis as soon as possible. Recovery is also related to the aspects of medical equipment supporting recovery processes, for example, recommendations of supportive bandages or medications (e.g., inflammation inhibitors).

Constraint-based systems can also be used for the recommendation rehabilitation exercises of a longer period of time. Examples thereof are marathon runners or tennis players restarting their exercises after a muscle injury or competitive skiers re-starting their training program after an injury. In all of these scenarios, time  constraints regarding the duration of exercise units (e.g, per week) including related obligatory training breaks and related resource constraints (e.g., in terms of the share of intensive exercises) have to be taken into account in order to avoid repeated injuries. An example of the application of constraint-based recommendation in the context of cardiac rehabilitation is presented in Dobrican et al. \cite{DobricanZampuieris2016RehabilitationConstraintBasedRiskAvoidance}. A related recommender tool generates physical exercises that fit a patients recommended training zones. Recommendations are determined on the basis of predefined constraints, for example, specifying duration and intensity of exercises. Furthermore, on the basis of constantly collected physical data, patients are guided and alerted in exceptional situations (e.g., cardiac arrhythmias).

\subsection{Fostering Healthy Living}\label{subsec:healthyliving}

For athletes as well as sports amateurs/enthusiasts, healthy living is crucial as it provides a solid foundation for success \cite{Abharietal2019,Bhimavarapuetal2021,Geetal2015,Yangetal2017}. Healthy living can have a positive impact on aspects such as physical fitness, faster recovery, increased performance,  increased success rates in competitions, and longevity, for example, in terms of being able to extend the duration of ones professional career. Healthy living entails a broad spectrum of  aspects such as healthy nutrition, hydration, sleep, stress management, and balanced lifetime. In this overview, we focus on the role of recommenders in supporting different aspects of healthy living.

\subsubsection{Nutrition}

To support the overall goal of avoiding an increase of body weight (leading to obesity in the worst case), there is a need of supporting healthy food decisions which is triggered by the increasing amount of different recipes which makes it time-consuming to make healthy food choices.  In this context, collaborative filtering can help to identify food items an athlete could be interested (depending on his her previous food preferences and further parameters such as physical condition) in \cite{Shrimaletal2021}. Collaborative filtering is specifically useful to find recipes and menus relevant for athletes in a similar situation in terms of personal goals, physical condition, and food preferences \cite{AlcarazHerrera2022,Donciuetal2011,Nietal2019}. Grouping athletes into similar categories this way also helps to better understand which recommendations can best help to motivate users to increase their personal performance.

\begin{example}[Recommendation of Sports Food] In the example in Table \ref{tab:healthfoodrecommendation} it is assumed that the user community, for example, tennis players, consume energy bars throughout competitions and training sessions. In order to generate serendipity in such scenarios, energy bars already consumed by nearest neighbors could be recommended to the current user -- the nearest neighbor of the current user is $u_1$, i.e., bar$_5$  (having a good rating) can be recommended to the current user. One step further (from the basic scenario) is to provide recommendations that help to improve the eating behavior of the current user with the objective of improving the personal sports performance \cite{Vandeputteetal2022}. 
\end{example}

\begin{table}[ht]
\centering \caption{Applying collaborative filtering (CF) in sports food (energy bars) recommendation where users $u_i$ provide ratings for different products on a scale $1$(low) .. $5$(high).}
\begin{tabular}{|c|c|c|c|c|c|c|c|c|c|c|c|c|c|c|c|c|c|c|c|c|c|c|c|c|c|} 
\hline
user    & bar$_1$   & bar$_2$ & bar$_3$ & bar$_4$ & bar$_5$ & bar$_6$  \tabularnewline \hline \hline
u$_1$   & 5.0       &     -   &     2.0 &   -     &   4.0   &   1.0    \tabularnewline \hline 
u$_2$   & -         &     -   &     5.0 &   2.0   &   -     &   4.0    \tabularnewline \hline 
u$_3$   & 4.0       &     4.0 &     -   &   2.0   &   5.0   &   4.0    \tabularnewline \hline \hline
current & 5.0       &     ?   &     1.0 &    ?    &    ?    &    ?     \tabularnewline \hline
\end{tabular} 
\label{tab:healthfoodrecommendation} 
\end{table}

Integrated food recommendation, for example, in terms of having meal plans for a whole week or even longer, requires the integration of constraint-based  recommendation technologies which help to assure that some eating-habits related constraints are satisfied. Examples of such constraints are: (1) the maximum amount of allowed calories per day must not be exceeded, (2) the share of carbohydrates and proteins should be in line with the requested amount of the envisioned sports activities for the same week, (3) sustainability aspects \cite{Spindler2023} need to be taken into account, and (4) the consumed food items should be as healthy as possible, evaluated, for example on a \emph{Nutri-Score} scale \cite{Majjodietal2022}. Such menu recommendation tasks are also relevant in the context of group decision making, think about the scenario of a menu planning for the whole week  which has to take into account the food preferences of all team members -- in this context, also \emph{fairness} aspects play a major role, i.e., in the long run to take into account the preferences of each group member in a nearly equal fashion \cite{Felfernig2018_1,Lietal2023}.

Espin et al. \cite{Espinetal2016} introduce an approach to recommend nutrition to elderly people. In this context, constraints take over the core recommendation role in terms of relating user preferences (including a users physical condition) into corresponding nutrition recommendations.  Such approaches need to be combined in a hybrid fashion with orthogonal approaches that help to achieve recommendation diversity. For example, in terms of carbohydrates, a person should not always receive a recommendation of spaghetti with tomato sauce but include a kind of variety in the recommended type of noodles, for example, instead of tomato sauce, use zucchini noodles with tomato sauce.

\subsubsection{Training Practices}

In the recent years, walking, hiking, and climbing have become an increasingly popular means of health promotion, not only for the elderly but rather for persons from all different ages \cite{Ivanovaetal2022,IvanovaIntroducingContext2023,IvanovaWald2023,Lietal2021,SasakiTakama2013,Takamaetal2015}. Walking is a good means to improve the physical condition. An important aspect in this context is to support people in identifying walking routes that require similar efforts compared to the currently preferred ones and thus relieving users from complex search tasks \cite{Takamaetal2015}. Related recommendation algorithms should take into account the aspect of diversity, for example, in terms of changing (but acceptable) difficulty levels as well as in terms changing routes taking into account a users preferences, for example, preferred mountain views, lakeside views, or in the other extreme, city walks. Such recommendations can be based on collaborative filtering which better support serendipity effects, content-based recommendation with a focus on identifying very similar routes, and knowledge-based (specifically critiquing-based \cite{Chen2012}) recommendation if the goal is to explore the set of possible routes. In this context, items (e.g., routes or training sessions) should be recommended in such a way that user engagement is maintained in the long run \cite{Berkovskyetal2010,Tragosetal2023} and specific physical conditions  (e.g., recommendation of physical practices for underweights \cite{JuliantBaizalDharayani2023}) and psychological aspects (e.g., personality-aware recommendations \cite{Khwajaetal2019}) of persons are taken into account.

\subsection{Event Handling \& Recommendation} \label{subsec:eventhandling}

\subsubsection{Sports Events}

Collaborative filtering (CF) can be applied for determining recommendations of sports events that might be of relevance for a person \cite{Achilleosetal2021}. CF recommenders support serendipity effects, i.e.,  a sports event recommender would not just focus on the current user preferences but potentially identify events of further interest not directly related to the current preferences. In contrast, content-based filtering would focus on very similar events \cite{Achilleosetal2021,Nguyenetal2019}, for example, a marathon runner who participated in marathons in New York and Boston up-to-now would receive related recommendations of future events, including, for example, other marathon events in the US. Case-based, more precisely, critiquing-based recommender systems \cite{Chen2012} can help to navigate and better understand the whole item space and -- if the corresponding data is available -- help to efficiently narrow down the option space, for example, only events should be recommended, where opponents participate who have a similar competitive level as the "current user". This scenario is relevant, for example, in the context of amateur tennis competitions where players prefer to participate in events with a  chance of  challenging but beatable opponents. 

\subsubsection{Event Designs}

In the context of event recommendation, there also exist applications for group recommender systems. Examples thereof are (1) a group of friends has to decide about one European league match to visit in the current season, (2) all members of a tennis team have to decide about the mode of the yearly team-internal championships, and (3) members of a soccer league management board have to decide about future changes in terms of the number of participating teams and the championship format -- for a related example, see Table \ref{tab:grouprecommendationsoccerleague}. In this example, following a simple \emph{majority-voting} preference aggregation, the preferences of managers 1 and 3 made it to the final recommendation. 

\begin{table}[ht]
\centering \caption{Group recommendation: deciding about a new format for a soccer league. Recommendations are determined based on \emph{majority voting}.}
\begin{tabular}{|c|c|c|c|c|c|c|c|c|c|c|c|c|c|c|c|c|c|c|c|c|c|c|c|c|c|} 
\hline
  decision               & \multicolumn{2}{c|}{\#teams} & \multicolumn{2}{c|}{playing mode (\#rounds)}          & \multicolumn{2}{c|}{\#relegation teams}     \tabularnewline \cline{2-7}
 makers  & 10 & 16                       & 4   & 2 qualification 2 relegation    & 1 & 2                                             \tabularnewline \hline \hline
manager$_1$      & x  & -                        & -         & x                               & x & -                                             \tabularnewline \hline 
manager$_2$      & -  & x                        & x         & -                               & - & x                                             \tabularnewline \hline 
manager$_3$      & x  & -                        & -         & x                               & x & -                                             \tabularnewline \hline  \hline
recommendation   & x  & -                        & -         & x                               & x & -                                             \tabularnewline \hline
\end{tabular} 
\label{tab:grouprecommendationsoccerleague} 
\end{table}

Beyond the exemplified basic group decisions, professional sports management develops consumer-centered strategies that take into account the influence of elite athletes (with a kind of star status) on customer demand \cite{ChmaitetalTennisSuperstars2020} which can be expressed, for example, in terms of sold tickets, sold merchandises, and media representation of a sports organization. In the context of large tennis events such as the Australian Open, a clear influence of individual players on stadium attendance could be observed \cite{ChmaitetalTennisSuperstars2020}. Having this aspect in mind, recommender systems can be applied to support the allocation of players to different tennis courts taking into account   related aspects such as star status of opponents, quality of games the opponents previously played against each other, and also emotional aspects, for example, the degree of rivalry among the opponents. Following this idea, sports recommender systems can be used to foster fan engagement and increase corresponding visitor retention rates.

\subsubsection{Event-related Items \& Games}

Basic event or team-related recommendation scenarios range from the merchandising of basic fan items (e.g., tshirts) to items such as sport awards associated with the winning of a  competition (event) \cite{Pichletal2018SportsAwards}. Recommenders can help to personalize the consumption of event-related items such as news and videos and thus also help to assure a sustainable engagement of a fan community.  Sports news and videos related to teams or players are often associated with strong emotions and opinions which has to be taken into account  \cite{Lenhart2016CombiningCA,Sanchezetal2012}. When commenting a sports event, commentators are completely overwhelmed by the vast amount of information available for individual players and teams. On the basis of knowledge about individual players currently in-focus and knowledge about information preferences of the audiences, a recommender system can help to propose relevant commentaries in real-time assuring to keep the audience interested.  Subramaniyaswamy et al. \cite{Subramaniyaswamy2018} introduce such a system in the context of commenting cricket games. Furthermore, sports event related recommendations can also support the selection of specific games to bet on -- for  details, we refer to De Pessemier et al. \cite{Pessemieretal2018}.

\subsection{Explanations} \label{subsec:motivation}

Explanations are relevant in different recommendation contexts and  play a crucial role in terms of achieving different goals \cite{FeelyetalExplainableRaceTimePredictions2020,Tintarev2011}. These goals will now be discussed in the context of sports-related recommendations.

\subsubsection{Explanation Goals}
 Explanations can have various  goals including (1) assuring trust in a recommendation, (2) increasing a users item domain knowledge, (3) persuade (motivate) a user to perform specific actions \cite{CoppensetalMotivationMove2023,Deepaketal2022,Majjodietal2022,Tragosetal2023,Vandeputteetal2022,Zhaoetal2020}, (4) assuring transparency of recommendations, and  (5) assuring efficiency, i.e., to support fast decisions -- for related details, we refer to Tintarev and Masthoff \cite{Tintarev2011}. Specifically in the context of group recommendation scenarios \cite{Felfernig2018_1}, further goals need to be taken into account, for example: (1) fairness (taking into account the preferences of all group members) and (2) consensus (establishing agreement among group members \cite{TranetalConsensusModelsUMUAI2023}. Thus, explanations can be regarded as  important  to assure the interpretability of recommendations \cite{Tintarev2011}. An overview of explanation goals in sports-related recommendation contexts if provided in Table \ref{tab:explanationgoalsinsports}.

\subsubsection{Motivation in Sports}

When it comes to the motivation of individuals or group members, different dimensions of persuasion play a major role \cite{Cialdini1993}. (1) \emph{recommendations of experts} \cite{MartinetalSmartloss2016Weightloss,YomTovetalEcouragingPhysicalActivityDiabetes2017} can be regarded as relevant since experts are typically have the relevant item domain knowledge (e.g., a tennis-related recommendation of Roger Federer has a high probability of being followed by tennis enthusiasts), (2) the \emph{opinions of friends and colleagues} (i.e., the social environment) typically has a major impact on a persons own decision (e.g., since my friends have already purchased specific running shoes, I am inclined to purchase the same), (3) creating sparsity of items can create urgency in decision situations and make users to take a decision earlier (the deadline for registering for the running event is in two days), (4) commitment can help to keep a goal in mind, i.e., users who confirm to do something in public have a lower probability of giving up (e.g., having announced the participation in marathon run forces a user to positively complete this process), (5) reciprocity is a major mechanism to trigger actions in situations where a user has received some productive help in the past (e.g., if someone helped  with improving my tennis in the past, I will be more willing to act as his/her training partner in the future), and (6) if you like some persons, for example, due to their positive behavior in the past, you will more easily take a related decision (e.g., you tend to join teams with similar opinions and attitudes of life). Using such persuasive techniques can have a huge impact on the way a user reacts on recommendations.  

Further related aspects are (1) \emph{need for completion} \cite{WangetalTaskCompletion2021} referring to the idea that users tend to  complete their online tasks, in other words, if some of the tasks are still open and marked with a "red flag", a user tends to make this green as well (e.g., if there is still one short run missing in the marathon preparation for this week and this is indicated with a corresponding red flag, this motivates users to change the situation correspondingly) and (2) \emph{comparison-related motivation} \cite{LockwoodPinkus2008} referring to the concept of being enthusiastic about comparing his/her own performance with the performance of opponents and try to improve \cite{Berkovskyetal2010}. Examples thereof are \emph{pace time} in running where runners try to improve their own pace compared to their current best or the pace of friends or opponents and \emph{itn number} in tennis leading to the same result. These numbers can be found in many types of sports and are always a major means to motivate for more competition and -- of course -- related training practices.

\begin{table}[!]
\centering  
\caption{Goal dimensions of explaining sports recommendations.} 
\begin{tabular}{|C{1.5cm}|C{2.5cm}|C{5.5cm}|} 
\hline
    dimension &  goal & example explanation \tabularnewline \hline \hline 
   trust                 &  increase trust in recommendation & \emph{runners of similar age and physical performance achieved a similar marathon time with the following training program} \tabularnewline \hline 
   domain knowledge      & increase sports-related domain knowledge & \emph{having breaks between intensive training sessions is crucial for regeneration and performance boost} \tabularnewline \hline 
   transparency          & make clear the reasons behind a recommendation &  \emph{player X and Y have been included in the team configurations for neutralizing opponent players U and V} \tabularnewline \hline 
   efficiency            & trigger fast decisions & \emph{you need to slow down immediately since you crossed the anaerobic threshold} \tabularnewline \hline
   persuasiveness        & trigger specific actions & \emph{you have communicated your participation in marathon X to your friends, i.e., continue your training sessions to achieve this goal} \tabularnewline \hline
   fairness        & create a feeling of fairness within a group  & \emph{tennis player X was assigned to the center court already twice -- this time, player Y will be assigned to the centre court} \tabularnewline \hline
   consensus       & help to achieve consensus within a group  & \emph{our most important goal is to increase the goal scoring rate -- to achieve this, player X is clearly the best option} \tabularnewline \hline
\end{tabular} 
\label{tab:explanationgoalsinsports} 
\end{table}

\section{Research Issues} \label{sec:researchissues}

 On the basis of our literature analysis, we have derived the following major open research issues regarding the application of recommenders in sports scenarios.

\emph{Recommendation of Bundles}. Many of the examples discussed in this article refer to basic recommendation scenarios, for example, the collaborative recommendation of running shoes. However, to exploit the full potential of recommendation technologies, bundle recommendation \cite{BaietalBundles2019,DengetalBundle2020} is needed in various situations. For example, the recommendation of running shoes could be combined with a recommendation of training pace times for the next marathon run integrated with explanations when it makes sense to dispose of the shoes and purchase new ones.

\emph{Evaluation Metrics}. Compared to traditional e-commerce applications of recommender systems, evaluation metrics in sports scenarios at least sometimes appear to be different. Which prediction quality is a satisfactory metrics when it comes to the selling of running shoes, new metrics are needed in recommendation scenarios such as deciding about the number of teams and gaming modes in a football league -- important related aspects are follow-up costs and estimated increased league quality which are both more related to the consequences of taking a specific decision.

\emph{Explanations}. Although known in theory, further related research is needed to better understand the best ways to integrate persuasive recommendations into sports-related recommendation scenarios. For example, a better understanding is needed in which way consequences of a choice should be explained, either in a positive fashion (following the pattern \emph{if you do X, you will achieve Y}) or negative fashion (following the pattern \emph{if you do not X, you will not achieve Y}). Other consequence explanations are going beyond basic motivation scenarios, for example, due to .. the economic value of hiring player X will  the  following ..). 

\emph{Psychological Aspects}. There are many aspects related to decision psychology which play a major role in the application of recommender systems. In general, decision biases are triggered by human decision short-cuts leading to potentially low-quality decisions. An example thereof are serial position effects referring to situations where items shown at the beginning and the end of a list a investigated more often and also chosen more often. For related details we refer to the work of Tran et al. \cite{AtasetalPreferenceConstruction2021}. A related aspect is also the so-called \emph{exercise paradox} which indicates that humans have hunter-gatherer genes through evaluation but at the same time we prioritize rest and inactivity due to the fact that food and energy are limited resources \cite{Smyth2019}. This situation triggers a need for systematic integration of motivation mechanisms into sports recommender systems. One related mechanism is the integration of (personalized) awards playing a central role in different gamification-based recommendation approaches \cite{Berkovskyetal2010,Zhaoetal2020}. Finally, a somehow open issue is to analyze in more detail which aspects have to be taken into account by music recommendation when applied in specific sports scenarios \cite{Chtourouetal2015}.

\emph{Realtime Recommendations.} An important aspect not covered by the majority of existing recommender systems are realtime functionalities which accompany athletes, for example, within the scope of training sessions. Examples of  realtime recommendations are: (1) recommendation of a new pacing   as an adaptation due to changed environmental conditions or performance issues due to rapidly changing temperatures and (2) a realtime recommender directly informing a tennis player how to adapt a swing to increase the stroke ground speed.

\emph{Large Language Models (LLMs).}  LLMs have excellent capabilities of understanding and generating human-like text sequences and thus have the potential to play a major role in the development of recommender systems \cite{DiPalma2023}. In this context, LLMs have to be investigated, for example, in terms of different approaches to extend user profiles, to make explanations more user-friendly, to improve the prediction quality of recommendations, and also to create more flexible user dialogs in the context of conversational recommendation \cite{Gao2021}. Specifically, sports recommender systems with corresponding LLM integrations have the potential to augment  sports-related knowledge, for example, in terms of new additional tactical options and training practices not taken into account up to now by a sports recommender \cite{connor2023large}. Examples of applying LLMs in different sports-related scenarios are depicted in Table \ref{tab:llmexamples}.

\begin{table}[!]
\centering  
\caption{Examples of applying large language models (LLMs) in sports recommendation (LLM answers have been shortended slightly).} 
\begin{tabular}{|C{1.5cm}|C{2.5cm}|C{6.5cm}|} 
\hline
    application scenario        &  LLM query  & LLM output \tabularnewline \hline \hline 
    recommendation of trainings &  Minimum duration of marathon preparation for an amateur? & \emph{The minimum duration for marathon preparation can vary depending on the individual's current fitness level, running experience, and goals. However, a common recommendation is to allow at least 16 to 20 weeks for marathon training, especially for beginners or amateurs.}\tabularnewline \hline
    tactical \& strategic planning &  How to play tennis against a moonball player? & \emph{Take the ball early and vary your shots.}\tabularnewline \hline
    recommending individuals \& teams & Would David Alaba be a good goalkeeper? & \emph{While Alaba is a highly skilled and adaptable player, transitioning to the goalkeeper position is a significant shift. Goalkeeping requires a unique set of skills, such as shot-stopping, positioning, handling crosses, and distribution, which are different from the skills needed for outfield positions.}\tabularnewline \hline
    handling injury \& illness & One tip how to avoid a tennis elbow? &  \emph{Using a grip that is too small or too large can contribute to increased stress on the forearm muscles and tendons. A grip that is too small may lead to increased wrist movement, while a grip that is too large can make it challenging to properly control the racket.} \tabularnewline \hline
    fostering healthy living \& illness & One tip what to eat before a marathon run?  & \emph{About 3 to 4 hours before the marathon, aim for a well-balanced meal that includes complex carbohydrates, lean protein, and a small amount of healthy fats.} \tabularnewline \hline
    event handling \& recommendation \& illness & One tip how to assign players to courts throughout a championship? & \emph{Matchups between regional rivals can add an extra layer of excitement to the viewing experience. For example, games between teams like "Sturm Graz" and "Wolfsberger AC" or any matchups involving local rivals in Vienna can be intense and captivating.} \tabularnewline \hline
\end{tabular} 
\label{tab:llmexamples} 
\end{table}

\section{Conclusions} \label{conclusions}
There is a general trend of applying recommender systems not only in the online world but in the "real world" of working environments, recreation, and leisure time activities. In this article, we  provide an overview of the state-of-the-art in applying recommender systems in sports-related scenarios. In terms of applied recommendation algorithms, the identified research contributions range from basic collaborative filtering and content-based filtering  to hybrid and group-based recommendation. To increase understandability, we provide a couple of working examples showing the application of recommender systems in sports scenarios. In order to stimulate future topic-related research, we have discussed a couple of related open research issues.

\section{Acknowledgements} This work is supported by Graz University of Technology.

%
%

\bibliographystyle{abbrv}


\bibliography{bibliography}

\end{document}